\documentclass[onecolumn,letterpaper,11pt]{article}

\usepackage{xcolor, bbm, lastpage, color, amsfonts, cite, gensymb, here}
\usepackage{url, array, amsmath, fancyhdr, verbatim, graphicx, microtype}
\usepackage{amssymb, mathtools, bm}
\usepackage[footnotesize]{caption}
\usepackage[tmargin=1in,bmargin=1in,lmargin=1in,rmargin=1in]{geometry}
\usepackage{IEEEtrantools}
\usepackage[inline]{enumitem}
\usepackage{booktabs}
\usepackage{soul}
\usepackage{wrapfig}
\usepackage{subcaption}

\usepackage{titlesec}

\titleformat*{\section}{\large\bfseries}

\titleformat*{\subsection}{\normalsize\bfseries}

\usepackage{setspace}

\title{\vspace{-3cm} \singlespacing Advances in Multi-agent Reinforcement Learning: \\ Persistent Autonomy and Robot Learning Lab Report 2024}
\author{Reza Azadeh \footnote{Persistent Autonomy and Robot Learning (PeARL) lab, Miner School of Computer and Information Sciences, University of Massachusetts Lowell, Lowell, 01854,  \texttt{reza\_azadeh@uml.edu}}}

\date{}

\begin{document}
\bstctlcite{IEEEexample:BSTcontrol}
\pagenumbering{arabic}

\maketitle

Multi-Agent Reinforcement Learning (MARL) approaches have emerged as popular solutions to address the general challenges of cooperation in multi-agent environments, where the success of achieving shared or individual goals critically depends on the coordination and collaboration between agents. However, existing cooperative MARL methods face several challenges intrinsic to multi-agent systems, such as the curse of dimensionality, non-stationarity, and the need for a global exploration strategy. Moreover, the presence of agents with constraints (e.g., limited battery life, restricted mobility) or distinct roles further exacerbates these challenges.


This document provides an overview of recent advances in Multi-Agent Reinforcement Learning (MARL) conducted at the Persistent Autonomy and Robot Learning (PeARL) lab at the University of Massachusetts Lowell. We briefly discuss various research directions and present a selection of approaches proposed in our most recent publications. For each proposed approach, we also highlight potential future directions to further advance the field.

\section{Integrating Relational Networks into Value-Based Factorization} 

Among the various MARL methods for cooperative tasks, the Centralized Training with Decentralized Execution (CTDE) paradigm has become a prominent approach for addressing numerous cooperative challenges, including but not limited to those previously mentioned. Although CTDE-based approaches have demonstrated state-of-the-art performance in tackling coordination tasks, they lack the ability to steer the coordination strategy to which the algorithms converge. This limitation arises because these approaches
\begin{enumerate*}[label=(\roman*)]
    \item assume agents within the team are identical and
    \item do not incorporate mechanisms to specify inter-agent relationships. 
\end{enumerate*}
To address this issue, some algorithms have adopted the concept of coordination graphs, enabling the sharing of action-values among agents through deep learning. However, these methods require learning to weigh and distribute action-values from a given undirected graph, which may not fully account for the influence of certain behaviors over others. Consequently, the problem of handling agents with diverse attributes, team structures, or prioritization among agents remains unresolved.

This work proposes a novel framework that leverages a relational network to capture the relative importance and priorities agents assign to one another, enabling the discovery of new cooperation strategies~\cite{findik2023impact}. Our framework, the Relationship-Aware Value Decomposition Network (RA-VDN), integrates the relational graph into the CTDE paradigm to address cooperative MARL challenges. Unlike the approach in~\cite{haeri2022reward}, our framework does not rely on reward sharing among agents based on relationships. Instead, it modifies how agents contribute to the team reward, as determined by the relational network. Ultimately, agents still receive the same individual reward provided by the environment.

\begin{figure}[h]
    \centering
    \includegraphics[width=0.99\textwidth]{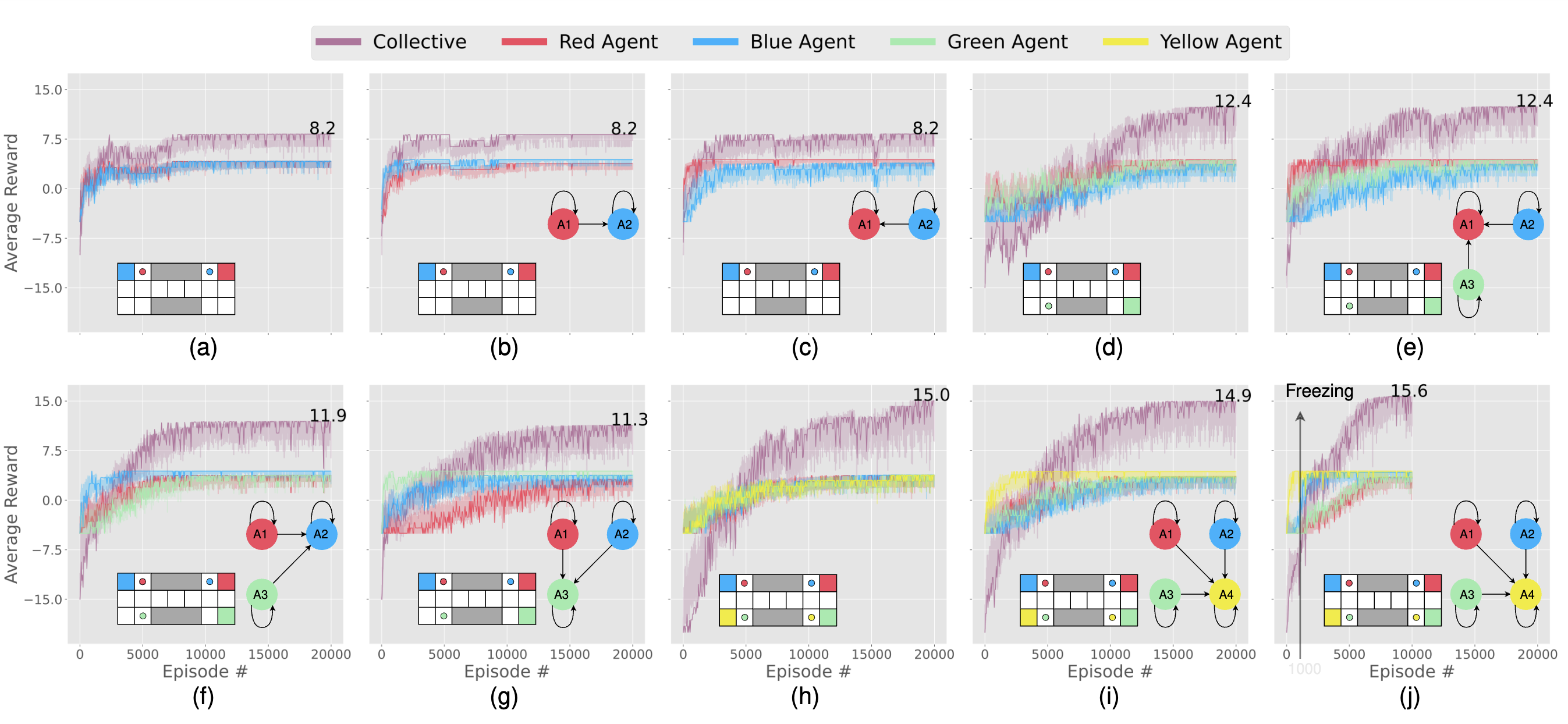}
    \caption{\emph{Switch} environment results with varying number of agents and different relational networks.}
    \label{fig:all_switch}
\end{figure}

We evaluated our approach through several experiments in two distinct environments, each involving a cooperative task among multiple agents. Our method was compared against the state-of-the-art Value Decomposition Networks (VDN) algorithm. Figure~\ref{fig:all_switch} presents the results of applying our algorithm to the \emph{switch} environment. The results and comparisons yield three main findings. First, unlike other methods, our approach is highly effective in guiding agents toward specific team behaviors dictated by the relational network. Second, the experiments demonstrate that our framework enables agents to discover new behaviors that facilitate successful task completion while preserving the team's essential structure. Lastly, our findings indicate that the proposed approach accelerates the learning of team behaviors, particularly in scenarios where agents face constraints and relational networks are critical. For more details, please refer to~\cite{findik2023impact}.

\subsection{Application to Malfunction Recovery in Multi-Robot Teams}
Multi-robot scenarios are common in various domains, including search and rescue operations, autonomous driving, and logistics and transportation. Coordination and cooperation among agents are essential for achieving the shared or individual goals of the mission. These factors become even more critical when addressing unexpected malfunctions, such as battery or motor failures~\cite{ahmadzadeh2014online, ahmadzadeh2014multi}. Thus, it is imperative for agents to cooperate effectively, recover from such failures, and adapt their strategies as a team to overcome these challenges.

\begin{figure}[h]
    \centering
    \includegraphics[width=0.5\textwidth]{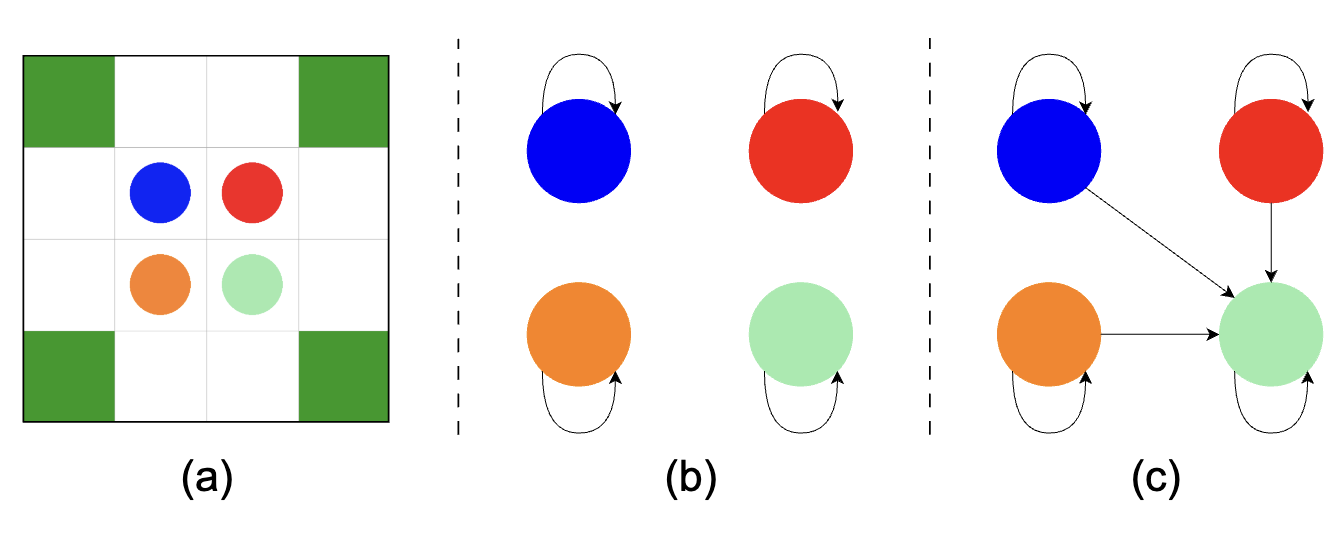}
    \caption{(a) multi-agent grid-world environment with four agents (circles) and four resources (dark green). The agent's action set includes moving in four directions as well as remaining idle. The agents can create the \emph{push behavior} when one agent remains idle and the other moves towards the idle agent; The Relational networks used by RA-VDN (b) before the malfunction and (c) after the green agent malfunctions.}
    \label{fig:malfunction_set}
\end{figure}
\begin{figure}[h]
    \centering
    \includegraphics[width=0.9\textwidth]{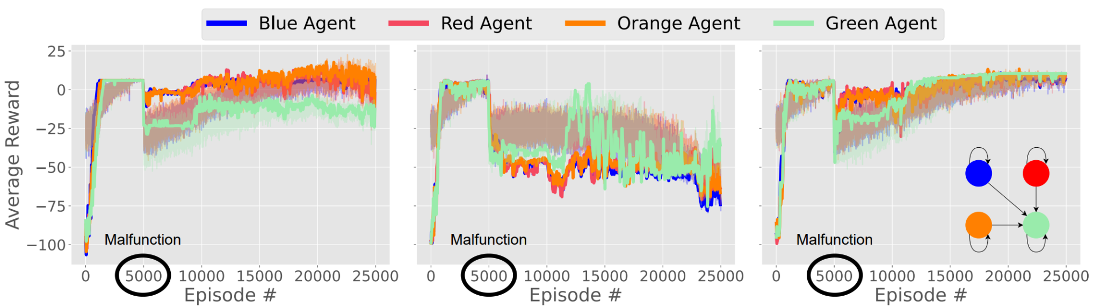}
    \caption{(left) Independent DQN, (middle) VDN, (right) RA-VDN. The malfunction happened at the 5000th episode.}
    \label{fig:malfunction_res}
\end{figure}

In this study, we analyze the application of our algorithm, RA-VDN, to the problem of malfunction recovery~\cite{findik2023collab}. As previously discussed, our algorithm leverages a relational network to capture the relative importance agents assign to one another, enabling faster adaptation to unexpected robot failures. To evaluate the effectiveness of our method, we conducted experiments in a multi-robot environment focused on a cooperative task with simulated random malfunctions. We compared our approach to the state-of-the-art Value Decomposition Networks (VDN) and Independent Deep Q-Networks (IDQN). We developed a simulated four-agent environment (Figure~\ref{fig:malfunction_set}) where the team was tasked with discovering a policy to complete a resource-gathering task, even when one team member became immobile. The results, shown in Figure~\ref{fig:malfunction_res}, demonstrate that our proposed approach facilitates effective cooperation within a multi-robot team, enabling faster adaptation to unforeseen malfunctions through the use of relational networks. 

\begin{figure}[b]
    \begin{minipage}[t]{.4\textwidth}
        \centering
        \raisebox{0.35\height}{\includegraphics[width=\textwidth]{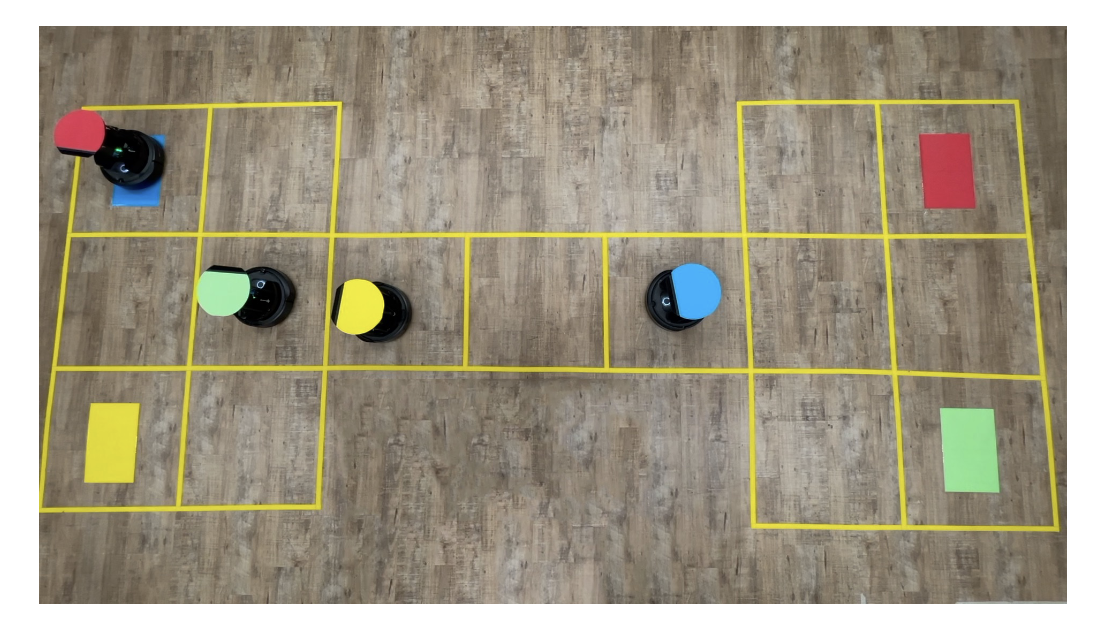}}
        \subcaption{\emph{Switch} environment with four agents (circles) and four stations (colored boxes) and a bridge.}\label{fig:real_switch}
    \end{minipage}
    \hfill
    \begin{minipage}[t]{.6\textwidth}
        \centering
        \includegraphics[width=\textwidth]{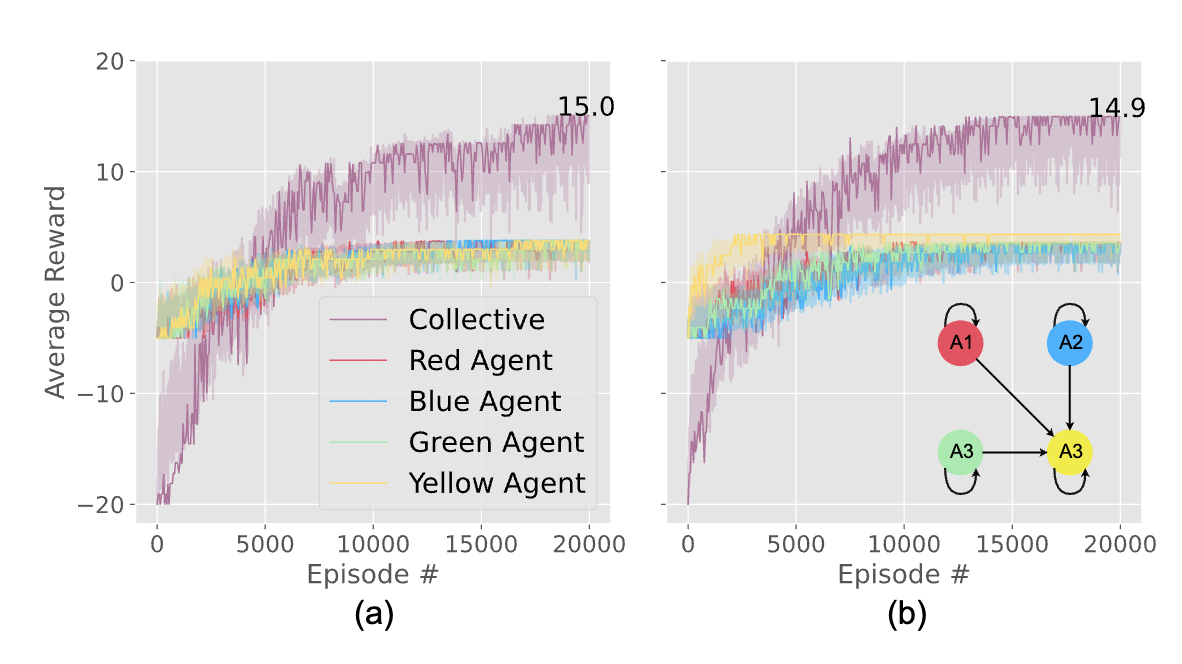}
        \subcaption{Results using (left) VDN (left) and RA-VDN (right) with the shown relational network.}\label{fig:4agent_switch}
    \end{minipage}  
    \label{fig:real_switch_result}
    \caption{Comparing VDN and RA-VDN in a four-agent cooperative scenario.}
\end{figure}

To validate the effectiveness of our proposed algorithm in a real-world environment, we replicated the Switch gridworld in a laboratory setting, as illustrated in Figure~\ref{fig:real_switch}. We utilized four Turtlebot4 mobile robots, each marked with a different color, mirroring the setup of our simulation experiments in the previous section. The results, shown in Figure~\ref{fig:4agent_switch}, compare our algorithm with VDN. For more details, please refer to~\cite{findik2023influence, findik2023collab}.

\clearpage
\section{Mixed Q-Functionals: Value-Based Methods in Continuous Action Domains}

As discussed in the previous section, the need for concurrent decision-making in discrete-action MARL environments introduces several challenges, such as non-stationarity, the curse of dimensionality, and the requirement for a global exploration strategy. In continuous-action MARL environments, these challenges become even more severe, to the extent that calculating each action's value becomes infeasible.

State-of-the-art continuous-action MARL approaches, such as Multi-Agent Proximal Policy Optimization (MAPPO) and Multi-Agent Deep Deterministic Policy Gradient (MADDPG), address these issues using policy-based methods. Similar to their single-agent counterparts, these algorithms are designed to optimize policy parameters for multiple interacting agents based on the expected return, thereby stabilizing the learning process. However, these methods suffer from sample inefficiency, particularly when compared to value-based methods.

Recent advances in innovative neural network architectures for single-agent RL domains have enabled the application of value-based concepts to continuous-action spaces. The resulting single-agent algorithm, known as Q-functionals, has been shown to outperform leading policy-based methods. Essentially, the Q-functionals approach transforms a state into a function that operates over the action space, facilitating the simultaneous evaluation of action-values ($Q(s,a)$) for multiple actions. This parallel processing capability enables efficient assessment of numerous actions per state, leading to more effective sampling in continuous environments for single-agent scenarios.

In this work, we introduced Mixed Q-Functionals (MQF), a novel value-based approach designed to address the sample inefficiency of policy-based methods for cooperative multi-agent tasks in continuous action spaces~\cite{findik2024mixed}. Unlike frameworks such as MADDPG, which rely on a policy network, MQF enables agents to utilize the concept of Q-Functionals—efficiently computing individual action-values—and combine them to enhance cooperation among multiple agents.

We evaluated our approach through six experiments across two distinct environments, each involving cooperative multi-agent tasks. Additionally, we introduced two baseline methods, Independent Q-Functionals and Centralized Q-Functionals, as novel value-based approaches for multi-agent tasks in continuous action domains. To the best of our knowledge, this work is the first to demonstrate the advantages of value-based methods over policy-based methods in cooperative MARL with continuous action spaces, opening new avenues for value-based strategies.

Our comparative analyses show that MQF consistently outperforms DDPG-based methods in similar settings, achieving optimal solutions and demonstrating faster convergence across six different scenarios

\subsection{Application to Robot Malfunction Recovery}

The application of multi-agent learning methods to complex robotics tasks has gained significant attention~\cite{kormushev2015robot}. MARL methods can discover and learn cooperative behaviors not only within multi-robot teams but also among different modules of a single robot (e.g., joints of a manipulator arm). The importance of such collaborative behaviors becomes particularly evident during unexpected robot malfunctions~\cite{ahmadzadeh2014multi}.

In this study, we aim to demonstrate the effectiveness of the Mixed Q-Functionals (MQF) algorithm in recovering from unforeseen malfunctions in robots with continuous action domains~\cite{findik2024relational}. We evaluated our algorithm through four experiments in a simulated multi-agent setting, focusing on malfunction handling in the ant robot environment. Our results indicate that our approach not only fosters more effective cooperation among agents (e.g., robot legs) but also accelerates adaptation to malfunctions by leveraging the capabilities of relational networks.

\begin{figure}[h]
    \centering
    \includegraphics[width=0.7\textwidth]{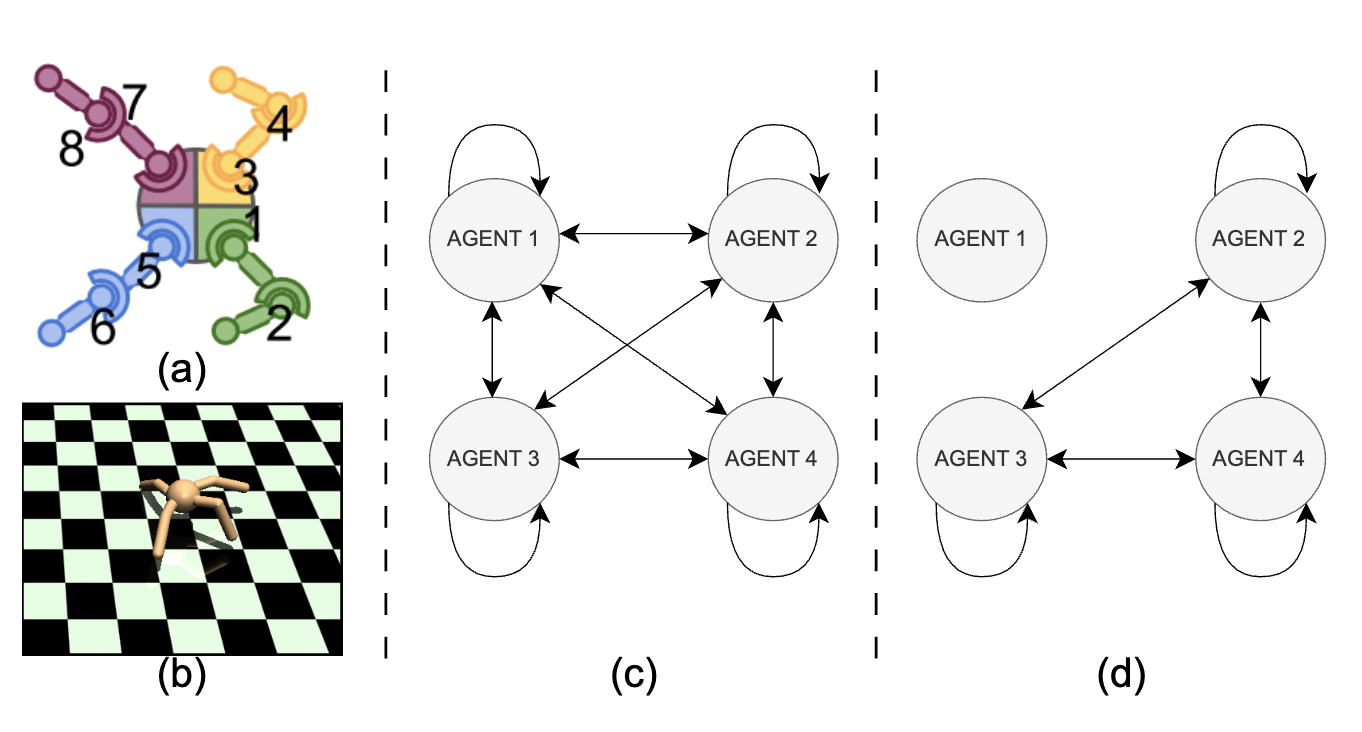}
    \caption{(a) Representation of a robotic ant featuring four agents, each distinguished by a different color and (b) The MaMuJoCo-Ant simulation environment. (c) Relational network used (c) before and (d) after agent 1 malfunctioned.}
    \label{fig:failure_set}
\end{figure}

\begin{figure}[h]
    \centering
    \includegraphics[width=\textwidth]{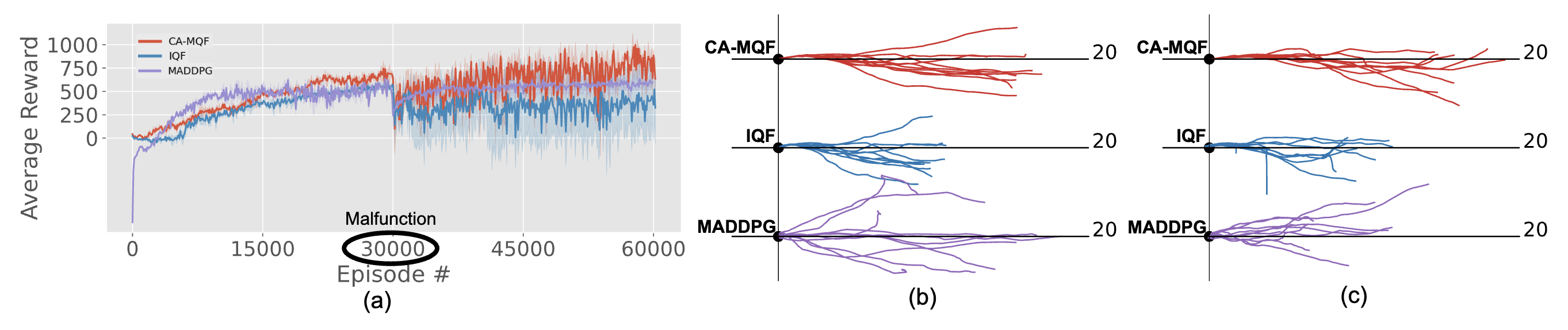}
    \caption{MaMuJoCo-Ant results: (a) Average team rewards before and after the malfunction at the 30000th episode. (b-c) Robot trajectories in the x-y plane: (b) before and (c) after malfunction, upon completing 30k and 60k training episodes, respectively. Compared to other methods, it can be seen that using MQF, the robot can cover more distance after malfunction occured.}
    \label{fig:failure_results}
\end{figure}

\section{Enhancing Multi-Agent Multi-Armed Bandit Team Performance through Relational Weight Optimization}

Another important aspect of enhancing cooperation in multi-agent systems—and consequently increasing team performance—is expediting the convergence to consensus. To study this problem, we consider Multi-Armed Bandits (MABs), a class of reinforcement learning problems that provide a simple model for simulating decision-making under uncertainty. In a MAB setting, an agent is presented with a set of arms (i.e., actions), each yielding a reward drawn from a probability distribution unknown to the agent. The agent's goal is to maximize its total reward, which requires balancing exploration and exploitation. Practical applications of MAB algorithms include news recommendations, online ad placement, dynamic pricing, and adaptive experimental design.

In contrast to single-agent scenarios, certain applications—such as search and rescue—require a team of agents to cooperate in order to maximize team performance. These problems are addressed using Multi-Agent Multi-Armed Bandit (MAMAB) algorithms.

Most existing algorithms rely on the presence of multiple agents and attempt to solve the problem using shared information among them while ignoring the relationships between team members. A few approaches, however, use graph representations and grouping to establish specific team structures. Representing team behavior with a graph ensures that the relationships between agents are maintained. However, existing works often fail to consider the weight of each relationship (graph edges) or require users to manually set these weights.

\begin{figure}[h]
    \centering
    \includegraphics[width=0.4\textwidth]{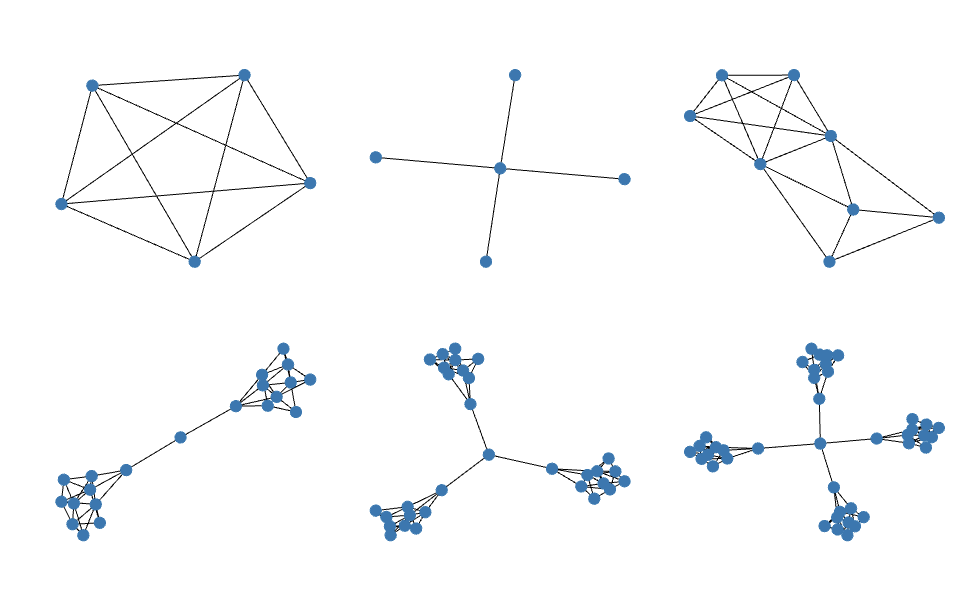}
    \caption{Graphical representation of the networks used in the experiments.}
    \label{fig:graphs}
\end{figure}

\begin{figure}[h]
    \centering
    \includegraphics[width=0.9\textwidth]{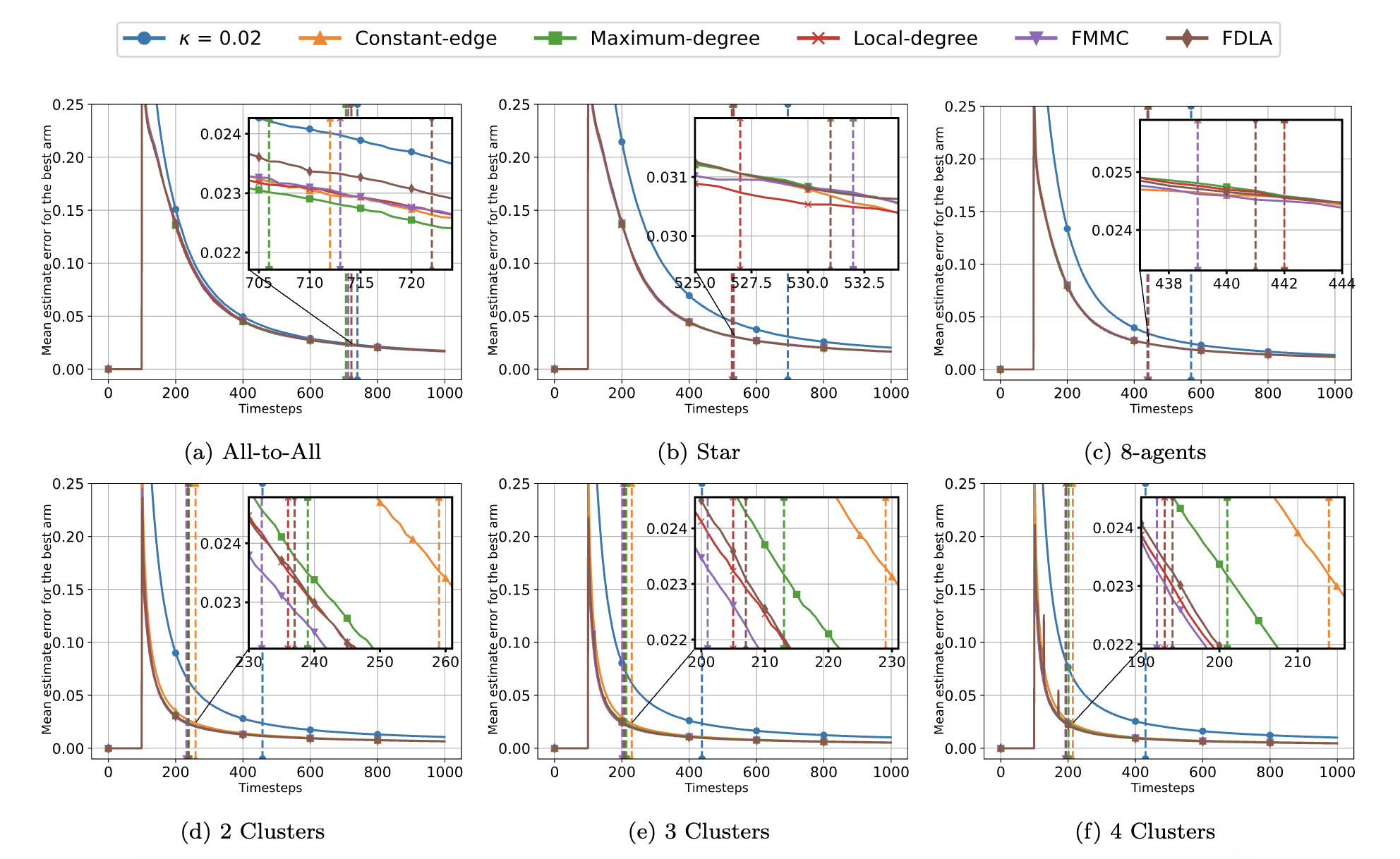}
    \caption{Comparison of team average of the errors between the estimated and true means of the best arm in different networks. The vertical dashed lines represent the time taken by the network to reach 5\% of the final value of the largest error among all algorithms.}
    \label{fig:graph_results}
\end{figure}

We introduced a new approach that combines graph optimization and MAMAB algorithms to enhance team performance by expediting the convergence to a consensus on arm means~\cite{kotturu2024relational}. While this work assumes the relational graph is given, the structure of the graph representing team behavior can also be inferred from data~\cite{kolli2024graph}. Our approach 
\begin{enumerate*}[label=(\roman*)]
    \item improves team performance by optimizing the edge weights in the graph representing the team structure in large constrained teams,
    \item does not require manual tuning of the graph weights,
    \item is independent of the MAMAB algorithm and only depends on the consensus formula, and
    \item formulates the problem as a convex optimization, which is computationally efficient for large teams.
\end{enumerate*}

We evaluated our method using the Coop-UCB2 MAMAB algorithm combined with six different graph optimization approaches in various team structures (Figure~\ref{fig:graphs}). Performance was measured by the time taken to converge to a consensus on the best arm mean. As shown in Figure~\ref{fig:graph_results}, our results indicate that the proposed method outperforms existing graph-based MAMAB algorithms with manual tuning or other adjustment heuristics in large, constrained teams. However, it has minimal impact on small networks. For more detail please see~\cite{kotturu2024relational}.

Future research directions include experimenting with real-world networks with different properties that may also benefit from optimized edge weights.

\bibliographystyle{IEEEtran}
\bibliography{references}

\begin{thebibliography}{10}
\providecommand{\url}[1]{#1}
\csname url@samestyle\endcsname
\providecommand{\newblock}{\relax}
\providecommand{\bibinfo}[2]{#2}
\providecommand{\BIBentrySTDinterwordspacing}{\spaceskip=0pt\relax}
\providecommand{\BIBentryALTinterwordstretchfactor}{4}
\providecommand{\BIBentryALTinterwordspacing}{\spaceskip=\fontdimen2\font plus
\BIBentryALTinterwordstretchfactor\fontdimen3\font minus \fontdimen4\font\relax}
\providecommand{\BIBforeignlanguage}[2]{{%
\expandafter\ifx\csname l@#1\endcsname\relax
\typeout{** WARNING: IEEEtran.bst: No hyphenation pattern has been}%
\typeout{** loaded for the language `#1'. Using the pattern for}%
\typeout{** the default language instead.}%
\else
\language=\csname l@#1\endcsname
\fi
#2}}
\providecommand{\BIBdecl}{\relax}
\BIBdecl

\bibitem{findik2023impact}
Y.~Findik, P.~Robinette, K.~Jerath, and S.~R. Ahmadzadeh, ``Impact of relational networks in multi-agent learning: A value-based factorization view,'' in \emph{2023 62nd IEEE Conference on Decision and Control (CDC)}.\hskip 1em plus 0.5em minus 0.4em\relax IEEE, 2023, pp. 4447--4454.

\bibitem{haeri2022reward}
H.~Haeri, R.~Ahmadzadeh, and K.~Jerath, ``Reward-sharing relational networks in multi-agent reinforcement learning as a framework for emergent behavior,'' \emph{arXiv preprint arXiv:2207.05886}, 2022.

\bibitem{ahmadzadeh2014online}
S.~R. Ahmadzadeh, M.~Leonetti, A.~Carrera, M.~Carreras, P.~Kormushev, and D.~G. Caldwell, ``Online discovery of auv control policies to overcome thruster failures,'' in \emph{2014 IEEE International Conference on Robotics and Automation (ICRA)}.\hskip 1em plus 0.5em minus 0.4em\relax IEEE, 2014, pp. 6522--6528.

\bibitem{ahmadzadeh2014multi}
S.~R. Ahmadzadeh, P.~Kormushev, and D.~G. Caldwell, ``Multi-objective reinforcement learning for auv thruster failure recovery,'' in \emph{2014 IEEE Symposium on Adaptive Dynamic Programming and Reinforcement Learning (ADPRL)}.\hskip 1em plus 0.5em minus 0.4em\relax IEEE, 2014, pp. 1--8.

\bibitem{findik2023collab}
Y.~Findik, P.~Robinette, K.~Jerath, and S.~R. Ahmadzadeh, ``Collaborative adaptation: Learning to recover from unforeseen malfunctions in multi-robot teams,'' in \emph{{MADG}ames workshop at {IEEE/RSJ} International Conference on Intelligent Robots and Systems ({IROS})}, 2023, pp. 1--6.

\bibitem{findik2023influence}
Y.~Findik, H.~Osooli, P.~Robinette, K.~Jerath, and S.~R. Ahmadzadeh, ``Influence of team interactions on multi-robot cooperation: A relational network perspective,'' in \emph{2023 International Symposium on Multi-Robot and Multi-Agent Systems (MRS)}.\hskip 1em plus 0.5em minus 0.4em\relax IEEE, 2023, pp. 50--56.

\bibitem{findik2024mixed}
Y.~Findik and S.~R. Ahmadzadeh, ``Mixed q-functionals: Advancing value-based methods in cooperative marl with continuous action domains,'' \emph{arXiv preprint arXiv:2402.07752}, 2024.

\bibitem{kormushev2015robot}
P.~Kormushev and S.~R. Ahmadzadeh, ``Robot learning for persistent autonomy,'' \emph{Handling Uncertainty and Networked Structure in Robot Control}, pp. 3--28, 2015.

\bibitem{findik2024relational}
Y.~Findik, P.~Robinette, K.~Jerath, and R.~Azadeh, ``Relational q-functionals: Multi-agent learning to recover from unforeseen robot malfunctions in continuous action domains,'' in \emph{2024 21st International Conference on Ubiquitous Robots (UR)}.\hskip 1em plus 0.5em minus 0.4em\relax IEEE, 2024, pp. 251--256.

\bibitem{kotturu2024relational}
M.~R. Kotturu, S.~V. Movahed, K.~Jerath, P.~Robinette, and R.~Azadeh, ``Relational weight optimization for enhancing team performance in multi-agent multi-armed bandits,'' in \emph{4th Modeling, Estimation and Control Conference ({MECC})}, 2024.

\bibitem{kolli2024graph}
A.~Kolli, R.~Azadeh, and K.~Jerath, ``Graph attention inference of network topology in multi-agent systems,'' in \emph{4th Modeling, Estimation and Control Conference ({MECC})}, 2024.

\end{thebibliography}

\end{document}